# PTD vs. PO effects in power and polarisation of PLANCK HFI – 100 beams

Fabio Noviello[a], Vladimir Yurchenko[a], Jean-Michel Lamarre[b], John Anthony Murphy[a]

[a] *National University of Ireland Maynooth, Maynooth, Co. Kildare, Ireland*

[b] *LERMA, Observatoire de Paris, 61 av. de l'Observatoire, 75014 Paris, France*

A method of accelerated computation of broadband far-off boresight patterns of telescope beams is extended and applied to the 100 GHz channels of Planck HFI (HFI – 100). Far-off boresight angle power patterns are computed with both GRASP-PTD and NUIM-PO software, considering cuts with an angular size of 10 degrees. The former reveal secondary oscillations of radial power distribution as compared with the latter. The difference for power patterns of a/b polarisation channels is also investigated.

## 1. Introduction

The necessity of controlling systematic errors in Planck HFI requires knowledge of the telescope beams over a large area comprising the far sidelobes. Physical optics (PO), together with the physical theory of diffraction (PTD) is the most adequate means of addressing this issue. This approach can be combined with geometrical optics (GO) and the geometrical theory of diffraction (GTD). Unfortunately, the computation time grows both with increasing frequency $v$ and boresight angle $\theta$. Another difficulty arises when computing broadband patterns. Namely, the number of single-frequency patterns needed for integration in order to reliably describe the desired broadband pattern.

## 2. Broadband integration.

Analytically solvable model beams emitted from circular apertures, having constant aperture fields modulated by Bessel functions, were studied in Ref. [1]. The essential point is that the spectral intensity is a double pseudo-periodic function in $k$ (wavenumber) as well as in $\theta$. We therefore propose substituting the integral over the spectral band at *each fixed spatial point $\theta$* with a band-equivalent integration over a suitable range of angles *around this point*. NUI Maynooth code has been developed to apply this technique to HFI beam patterns. The single-frequency beam pattern in Fig. 1 (left panel) was computed with the TICRA GRASP8 software package with a combined PO/PTD + GO/GTD approach. The horn aperture field was derived with a scattering matrix mode-matching method (NUI Maynooth SCATTER software). The right panel depicts the outcome of the application of a first version of the $\theta$-integration method (constant integration range for every $\theta$-point).

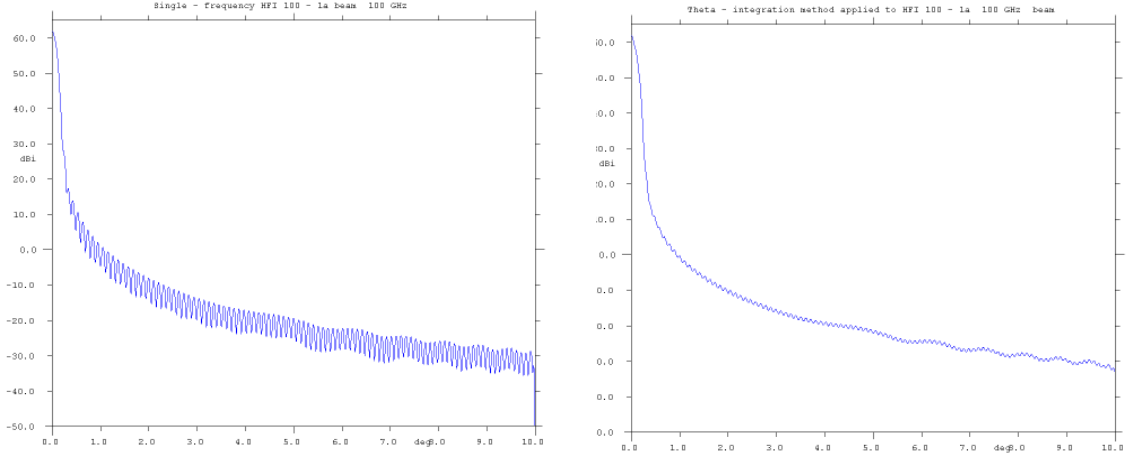

**Figure 1:** The left panel depicts a 10-degree radius beam pattern cut relative to the HFI 100-1a channel (100 GHz). On the right, the result of the simplest version of the $\theta$-integration method. is shown. The amplitudes of the far-off sidelobes are dampened as expected in broadband integration.

An interesting feature visible in Fig. 1 is the presence of secondary large-scale oscillations modulating the main far sidelobes. These are present in GRASP8 PO/PTD + GO/GTD simulations but not in PO-only computations run with NUI Maynooth fast PO code (not shown).

In order to characterise the spectral response of an antenna at least three frequencies are needed . In the case of HFI 100-1a we have chosen the central frequency together with with two others close to opposite band-edges (the bandwith being $\cong$ 30%). In Fig. 2 we compare the results of direct integration as opposed to the (normalised) sum of $\theta$-integrated cuts.

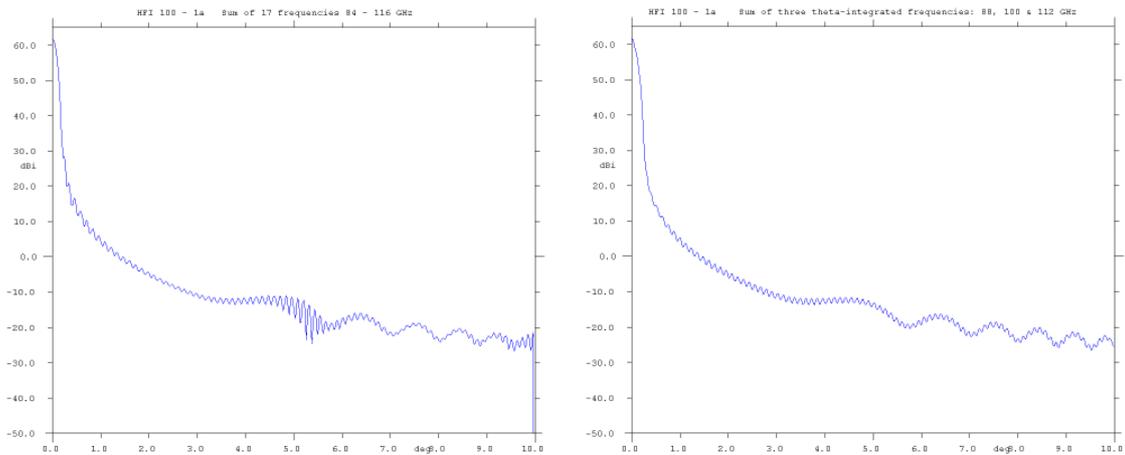

**Figure 2:** On the left we show a broadband HFI 100-1a cut obtained with the direct integration of 17 different frequencies. The right – hand side panel shows the outcome of the sum of 3 $\theta$-integrated cuts. There is a remarkable similarity.

The main advantage of our method is in a subtantial reduction of both computing time and necessary memory storage space for broadband beam patterns. Additional work is being carried out at NUI Maynooth both in refining the technique as well as extending it into the two-dimensional domain. Also, the origin of the secondary oscillations is currently under investigation.

**3. Polarisation.**

Since CMB polarisation anisotropies are at a level equal or below 10% of temperature anisotropies a careful characterisation of the polarisation properties of HFI beam patterns is required [2], [3]. We have calculated HFI 100-1a – 1b power patterns with both NUI Maynooth fast PO code and GRASP8 PO/PTD + GO/GTD approaches. These two approaches give consistent results for small boresight angles. Fig. 3 illustrates these results

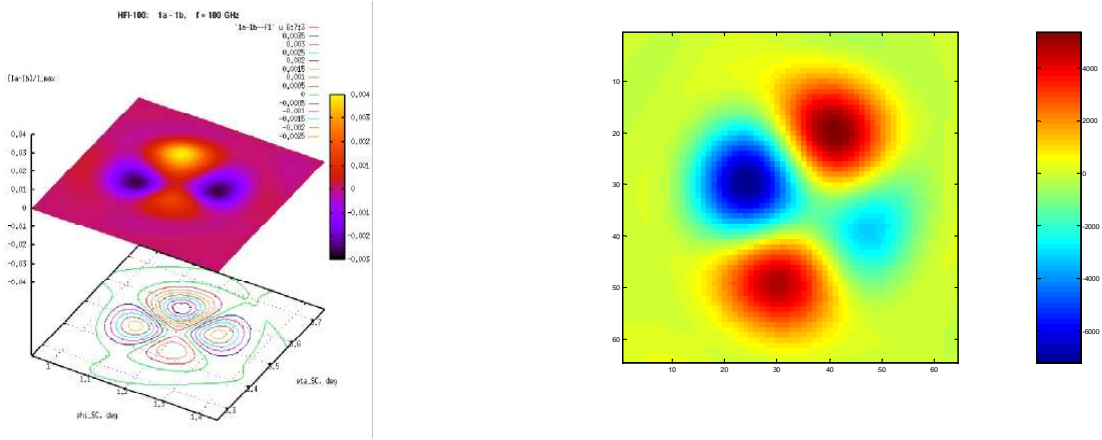

**Figure 3:** The left panel depicts HFI 100 1a-1b beam power pattern difference computed with NUI Maynooth fast PO code. Denoting with $Ix$ the peak intensity of any one of the two channels, we find that the ratio $(Ia - Ib)/Ia \cong 0.04$. The field of view is 0.8 deg$^{2.}$ On the right we have the same pattern computed with GRASP 8 in the PO/PTD + GO/GTD mode. In this case also $(Ia - Ib)/Ia \cong 0.04$.

**4. Conclusions**

The $\theta$-integration method offers substantial advantages over direct integration of broadband beam patterns both in computation time and in necessary storage space in computer memory. NUI Maynooth code has been developed to address this issue and its functionality is currently being extended. A GRASP 8 PO/PTD + GO/GTD approach shows secondary oscillations in HFI 100 –1 beam patterns for large boresight angles. Conversely, this approach gives consistent results with NUI Maynooth fast PO code for the main beams in the case of the a-b polarisation channels' power pattern difference.

.